\documentclass[aps,prb,twocolumn,showpacs,preprintnumbers,amsmath,amssymb,superscriptaddress]{revtex4}%

\usepackage{graphicx}
\usepackage{dcolumn}
\usepackage{bm}

\begin{document}
\preprint{PREPRINT (\today)}
%
\title{Site-selective oxygen isotope effect on the magnetic field
penetration depth in underdoped
Y$_{0.6}$Pr$_{0.4}$Ba$_2$Cu$_3$O$_{7-\delta}$ }
\author{R.~Khasanov}
\email{rustem.khasanov@psi.ch} \affiliation{Physik-Institut der
Universit\"{a}t Z\"{u}rich, Winterthurerstrasse 190, CH-8057,
Switzerland} \affiliation{Laboratory for Muon Spin Spectroscopy,
Paul Scherrer Institut, CH-5232 Villigen PSI, Switzerland}
\author{A.~Shengelaya}
\affiliation{Physik-Institut der Universit\"{a}t Z\"{u}rich,
Winterthurerstrasse 190, CH-8057, Switzerland}
\author{E.~Morenzoni}
\affiliation{Laboratory for Muon Spin Spectroscopy, Paul Scherrer
Institut, CH-5232 Villigen PSI, Switzerland}
\author{M.~Angst}
\affiliation{Physik-Institut der Universit\"{a}t Z\"{u}rich,
Winterthurerstrasse 190, CH-8057, Switzerland}
\author{K.~Conder}
\affiliation{Laboratory for Neutron Scattering, ETH Z\"urich and
Paul Scherrer Institut, CH-5232 Villigen PSI, Switzerland}
\author{I.~M.~Savi\'c }
\affiliation{Faculty of Physics, University of Belgrade, 11001
Belgrade, Serbia and Montenegro}
\author{D.~Lampakis}
\affiliation{Department of Physics National Technical University
Athens 157 80, Greece}
\author{E.~Liarokapis}
\affiliation{Department of Physics National Technical University
Athens 157 80, Greece}
\author{A.~Tatsi}
\affiliation{Department of Physics National Technical University
Athens 157 80, Greece}
\author{H.~Keller}
\affiliation{Physik-Institut der Universit\"{a}t Z\"{u}rich,
Winterthurerstrasse 190, CH-8057, Switzerland}
%

\begin{abstract}
We report site-selective oxygen isotope ($^{16}$O/$^{18}$O) effect
(OIE) measurements on the in-plane penetration depth $\lambda_{ab}
$ in underdoped Y$_{0.6}$Pr$_{0.4}$Ba$_2$Cu$_3$O$_{7-\delta}$,
using the muon-spin rotation ($\mu$SR) technique. A pronounced OIE
on the transition temperature $T_c$ as well as on
$\lambda_{ab}^{-2}(0)$ was observed, which mainly arises from the
oxygen sites within the superconducting CuO$_2$ planes (100 \%
within error bar). The values of the corresponding relative
isotope shifts were found to be $T_c/T_c= -3.7(4)$~\% and
$\Delta\lambda^{-2}_{ab}(0)/\lambda^{-2}_{ab}(0)= -6.2(1.0)\%$.
Our results imply that in this compound the phonon modes involving
the movement of planar oxygen are dominantly coupled to the
electrons.
\end{abstract}
\maketitle
%

More than $15$ years after the  discovery of the cuprate high
$T_c$ superconductors (HTSC), their pairing mechanism is still not
understood. For a long time, a role of phonons in this mechanism
was dismissed, despite evidence of their relevance, for example by
isotope experiments (see,  e.g.,
Refs.~\onlinecite{Franck94,Zech94,Zhao97,Hofer00,Khasanov03,Khasanov03b}).
One of the reasons for this was the prevailing belief that phonons
simply could not account for such high transition temperatures
($T_c$). The recent discovery of MgB$_2$, \cite{Nagamatsu01} where
phonons are clearly the pairing glue leading to an unexpectedly
high $T_c \approx 40\,{\mathrm{K}}$, clearly falsified this
belief. Many other widely cited reasons as to why phonons ought to
play no role in HTSC also do not stand to scrutiny.
\cite{Allen01,Kulic00} Recent work by inelastic neutron scattering
(INS) \cite{Mcqueeney99,Chung03} and angle resolved photoemission
spectroscopy (ARPES) \cite{Bogdanov00,Lanzara01} gave good
indications of a strong electron-phonon coupling in various HTSC,
invigorating the debate about the relevance of phonons.
\cite{Bussmann-Holder01,Eschrig03}

For conventional superconductors (including doped fullerens
\cite{Ramirez92} and MgB$_2$ \cite{Budko01}), key experimental
evidence for a phonon mediated pairing was provided by
measurements of the isotope effect (IE) on $T_c$. In complex
materials, such as the HTSC, $T_c$ may depend on many parameters
in a non-trivial way. Important complimentary information can be
obtained from the IE on the magnetic penetration depth $\lambda$.
Previous studies \cite{Zhao97,Hofer00,Khasanov03,Khasanov03b}
showed substantial oxygen isotope ($^{16}$O/$^{18}$O) effect (OIE)
on the in-plane penetration depth $\lambda _{ab}$, which indicates
an unusual (e.g., non-adiabatic) coupling of the electrons to
phonon modes involving the movement of the isotope substituted
atoms. It is important to identify the relevant phonon modes
responsible for this effect. Keen insight in this respect can be
obtained from studying the site-selective oxygen isotope effect
(SOIE), as we will detail below. While two reliable reports on the
SOIE on $T_c$ exist, \cite{Zech94,Zhao96} the SOIE on $\lambda
_{ab}$ was previously deduced only in a very indirect way from
magnetization measurements in the vicinity of $T_c$. \cite{Zhao96}

In this communication, we report on the SOIE on $T_c$ and on
$\lambda _{ab}$ in underdoped
Y$_{0.6}$Pr$_{0.4}$Ba$_2$Cu$_3$O$_{7-\delta}$, measured directly
and accurately by using the transverse field muon spin rotation
(TF-$\mu$SR) technique. \cite{Zimmermann95} The result clearly
indicates a strong coupling of the electronic subsystem to phonon
modes involving movements of the oxygen atoms in the CuO$_2$
plane, while suggesting that modes involving apical and chain
oxygen are less strongly coupled to the electrons. This agrees
with the conclusions observed from the combination of INS
\cite{Mcqueeney99,Chung03} and ARPES \cite{Bogdanov00,Lanzara01}
results, which were obtained in a completely different way and in
other HTSC.

Polycrystalline samples of
Y$_{0.6}$Pr$_{0.4}$Ba$_2$Cu$_3$O$_{7-\delta}$ were prepared by
standard solid state reaction. \cite{Conder01} In order to prepare
oxygen site-selective samples a two-step exchange process was
applied. \cite{Zech94,Conder01} In the first step [500$^{\rm o}$C,
35~h at 1.2~bar in $^{16}$O$_2$ ($^{18}$O$_2$) gas] two pairs of
completely oxygen substituted samples ($^{16}$O$_{\rm pac}$ and
$^{18}$O$_{\rm pac}$) were prepared [here indexes $p$, $a$ and $c$
mean planar (within CuO$_2$ planes), apical and chain oxygen,
respectively]. In the second step (330$^{\rm o}$C, 150~h, 1.2~bar)
two site-selective oxygen samples ($^{16}$O$_{\rm
p}$$^{18}$O$_{\rm ac}$ and $^{18}$O$_{\rm p}$$^{16}$O$_{\rm ac}$)
were prepared via annealing one $^{16}$O$_{\rm pac}$ sample in a
$^{18}$O$_2$ atmosphere and one $^{18}$O$_{\rm pac}$ sample in
$^{16}$O$_2$ gas. The other two samples (one $^{16}$O$_{\rm pac}$
and one $^{18}$O$_{\rm pac}$) were simultaneously annealed in the
same atmosphere as before. For the totally exchanged sample
($^{18}$O$_{\rm pac}$) the $^{18}$O content, estimated from the
change of the sample weight after annealing, was found to be
90(2)~\%. The oxygen-tracer diffusion calculation \cite{Conder01}
for the material used in this work (average measured grain size is
16~$\mu$m) shows that after annealing at 330$^{\rm o}$C, during
150~h more than 90~\% of the apex+chain sites and only about
10-15~\% of the plane sites should be exchanged. Keeping in mind
that there are 4 planar and 3 apical and chain oxygen sites in the
unit cell of Y$_{0.6}$Pr$_{0.4}$Ba$_2$Cu$_3$O$_{7}$, for the above
mentioned annealing procedure, one would expect about 45~\% of an
average oxygen substitution. The experimental estimate (from the
change of the sample weight) gives 41(2)~\% for both
site-selective samples, in good agreement with the theoretical
calculation.

\begin{figure}[htb]
\includegraphics[width=0.95\linewidth]{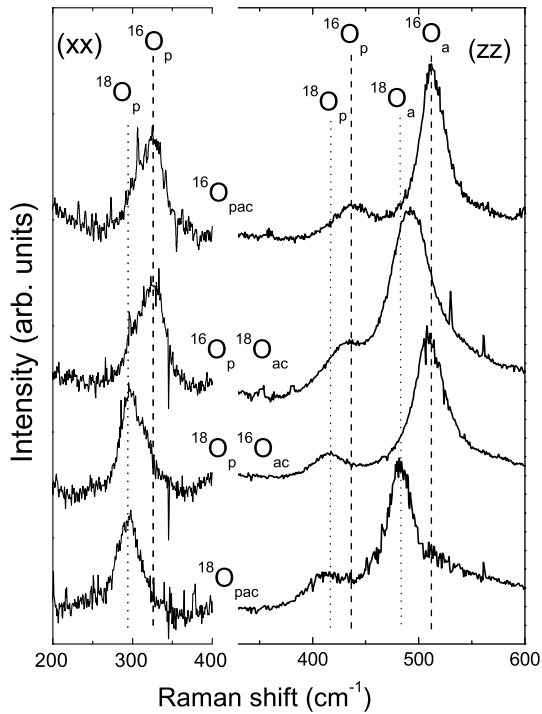}
\vspace{-0.5cm}
\caption{Room-temperature Raman spectra of the
Y$_{0.6}$Pr$_{0.4}$Ba$_2$Cu$_3$O$_{7-\delta}$ samples. For $xx$
polarization the line corresponds to the out-of-phase motion of
the planar oxygen atoms ($^{16}$O, 325~cm$^{-1}$; $^{18}$O,
294~cm~$^{-1}$). For $zz$ polarization the two lines correspond to
the in-phase motion of the CuO$_2$ plane oxygen ($^{16}$O,
436~$\mathrm{cm}^{-1}$; $^{18}$O, 415~$\mathrm{cm}^{-1}$) and to
the bond-stretching mode of the apical oxygen ($^{16}$O,
512~$\mathrm{cm}^{-1}$; $^{18}$O, 482~$\mathrm{cm}^{-1}$).  }
\label{Raman}
\end{figure}

The site-selectivity of the oxygen exchange was then checked by
Raman spectroscopy. The 488~nm and 514.5~nm lines of an Ar laser
were used for excitation. Figure \ref{Raman} shows Raman spectra
of all samples with $zz$ and $xx$ polarizations. In the $zz$
polarization for the  $^{16}$O$_{\rm pac}$ sample we observe two
lines, one corresponding to the in-phase motion along the $c$-axis
of the planar oxygen atoms ($436$~cm$^{-1}$) and one corresponding
to the bond-stretching mode of apical oxygen, which in this
compound is located at 512~cm$^{-1}$. \cite{Liarokapis03} In the
$xx$ polarization a line at $325$~cm$^{-1}$ is seen, corresponding
to the out-of-phase vibrations of the planar oxygen atoms.
\cite{Liarokapis03} In the $^{18}$O$_{\rm pac}$ sample, the Raman
lines are all shifted to lower frequencies with amounts agreeing
well with the ones obtained in Ref.~\onlinecite{Zech94},
indicating a nearly complete exchange of $^{16}$O with $^{18}$O.
In the site-selective sample $^{16}$O$_{\rm p}$$^{18}$O$_{\rm
ac}$, only the position of the apical oxygen line is shifted to
the lower frequency, whereas the lines corresponding to the plane
oxygen stay the same [apart from a small shift of one Raman line
(433~cm$^{-1}$ instead of 436~cm$^{-1}$), probably due to a small
unintentional partial substitution by $^{18}$O]. Note that the
apical line (492~cm$^{-1}$) is also slightly shifted from the
expected 482~cm$^{-1}$, indicating that oxygen exchange for the
apical and chain oxygen is slightly incomplete. In the
$^{18}$O$_{\rm p}$$^{16}$O$_{\rm ac}$ sample only two planar lines
are shifted, while the apical line stays the same. Therefore, the
Raman results confirm the almost complete site-selective oxygen
substitution in all samples.

\begin{figure}[tb]
\includegraphics[width=0.95\linewidth]{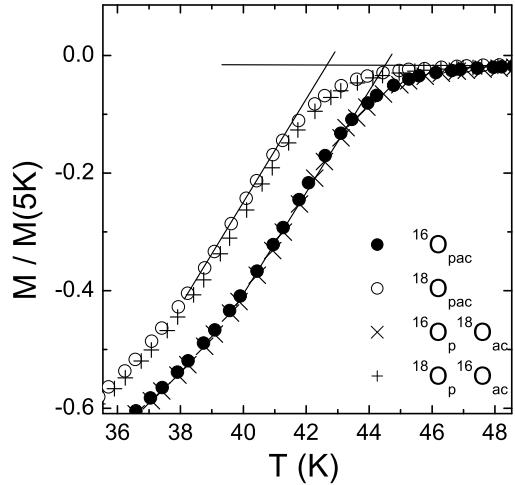}
\vspace{-0.5cm}
\caption{Section near $T_c$ of the normalized (to the value at
$5\,{\mathrm{K}}$) magnetization curves (1~mT, FC) of the
site-selective Y$_{0.6}$Pr$_{0.4}$Ba$_2$Cu$_3$O$_{7-\delta}$
samples. }
\label{Tc}
\end{figure}

To determine the site-selective oxygen isotope shift on $T_c$, low
field [1mT, field cooled (FC)] SQUID magnetization ($M$)
measurements were performed. Figure \ref{Tc} shows the $M(T)$
dependencies in the vicinity of $T_c$ for all four samples. It can
be seen that the $^{16}$O$_{\rm pac}$ sample has a higher $T_c$
than the $^{18}$O$_{\rm pac}$ one with nearly the same transition
width. The isotope shift $\Delta T_c = - 1.66(17)$~K for
completely substituted samples ($^{16}$O$_{\rm pac}$ and
$^{18}$O$_{\rm pac}$) is in agreement with previous results
\cite{Khasanov03,Franck91} ($T_c$ was extracted from the intercept
of the linear extrapolation as shown in Fig.~\ref{Tc}). The data
points of the site-selective $^{16}$O$_{\rm p}$$^{18}$O$_{\rm ac}$
sample coincide within error with those of the $^{16}$O$_{\rm
pac}$ sample, whereas the $^{18}$O$_{\rm p}$$^{16}$O$_{\rm ac}$
data points coincide with the $^{18}$O$_{\rm pac}$ ones,
indicating that the OIE on $T_c$ is related to the oxygen within
the CuO$_2$ planes. The results of the SOIE on $T_c$ are
summarized in Table~\ref{LambdaSSOIE}. The OIE exponent on $T_c$
was determined as $\alpha_{\rm O} = - d\ln T_c/d\ln M_{\rm O}$
($M_{\rm O}$ is the mass of the oxygen isotope).

In order to determine the SOIE on $\lambda_{ab}$ we performed
TF-$\mu$SR measurements at the Paul Scherrer Institute (PSI),
Switzerland, using the $\pi$M3 beam line. The samples were cooled
from far above $T_{c}$ in a magnetic field of 200~mT. In a powder
sample the magnetic penetration depth $\lambda$ can be extracted
from the muon-spin depolarization rate $\sigma(T) \propto
1/\lambda^{2}(T)$, which probes the second moment $\langle \Delta
B^{2}\rangle^{1/2}$ of the probability distribution of the local
magnetic field function $p(B)$ in the mixed state.
\cite{Zimmermann95} For highly anisotropic layered superconductors
(like the cuprate superconductors) $\lambda$ is mainly
determinated by the in-plane penetration depth $\lambda_{ab}$ : $
\sigma(T) \propto 1/\lambda_{ab}^{2}(T)$. \cite{Zimmermann95}
\begin{figure}[tb]
\includegraphics[width=0.95\linewidth]{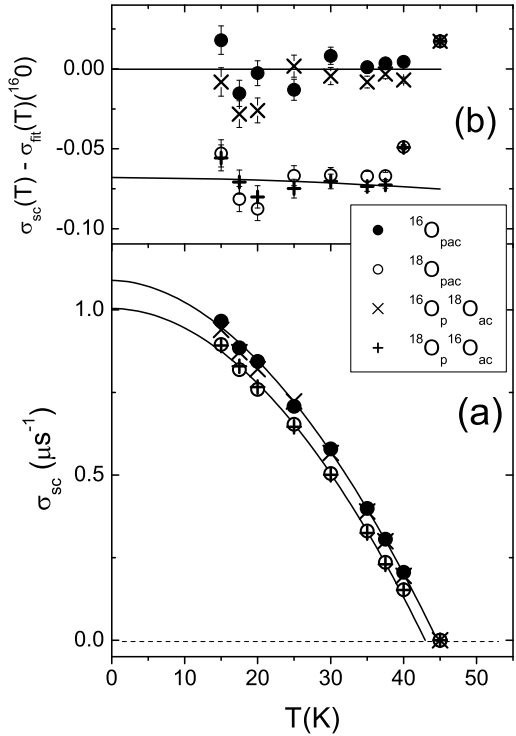}
\vspace{-0.5cm}
\caption{ (a) Temperature dependence of the depolarization rate
$\sigma_{sc}$ in site-selective
Y$_{0.6}$Pr$_{0.4}$Ba$_2$Cu$_3$O$_{7-\delta}$ samples  (200~mT,
FC). Data points below 10~K are not shown (see text for an
explanation). The solid lines correspond to fits to the power law
$\sigma_{sc}(T)/\sigma_{sc}(0)=1- (T/T_{c})^n$ for the
$^{16}$O$_{\rm pac}$ and $^{18}$O$_{\rm pac}$ samples. (b) Same as
in (a), but after subtracting the fitted curve  for the
$^{16}$O$_{\rm pac}$ sample. }
\label{sigma}
\end{figure}
The depolarization rate $\sigma$ was extracted from the $\mu$SR
time spectra using a Gaussian relaxation function $R(t) =
\exp[-\sigma^{2}t^{2}/2]$. As discussed in
Ref.~\onlinecite{Khasanov03}, the measured total depolarization
rate $\sigma$ is determined by three terms: a superconducting
($\sigma_{sc}$), an antiferromagnetic ($\sigma_{afm}$), and a
small nuclear magnetic dipole ($\sigma_{nm}$) contribution. The
antiferromagnetic contribution $\sigma_{afm}$ is present {\em
only} at low temperatures. Therefore data points below 10~K were
excluded in the analysis. The dipole contribution $\sigma_{nm}$ is
practically temperature independent below $100\,{\mathrm{K}}$, and
$\sigma_{sc}$ was determined by subtracting $\sigma_{nm}$ measured
above $T_{c}$ from $\sigma$ according to:
$\sigma_{sc}=\sqrt{\sigma^2-\sigma_{nm}^2}$. Figure \ref{sigma}
shows the temperature dependence of $\sigma_{sc}$ for the
Y$_{0.6}$Pr$_{0.4}$Ba$_2$Cu$_3$O$_{7-\delta}$ site-selective
samples.  It is evident that a remarkable oxygen isotope shift of
$T_c$ as well as of $\sigma_{sc}$ is present. More importantly,
the data points of the site-selective $^{16}$O$_{\rm
p}$$^{18}$O$_{\rm ac}$($^{18}$O$_{\rm p}$$^{16}$O$_{\rm ac}$)
samples coincide with those of the $^{16}$O$_{\rm
pac}$($^{18}$O$_{\rm pac}$) samples.
In order to put into evidence these results, the curve fitting the
temperature dependence of $\sigma_{sc}(T)$ for $^{16}$O$_{\rm
pac}$ (shown below) was subtracted from the experimental data
[Fig.~\ref{sigma}(b)]. It can be seen that the experimental points
for the two pairs of samples mentioned above coincide within the
error bars (nearly within the error bars in the case of
$^{16}$O$_{\rm pac}$ and $^{16}$O$_{\rm p}$$^{18}$O$_{\rm ac}$).
The data in Fig.~\ref{sigma} were fitted to the power law
$\sigma_{sc}(T)/\sigma_{sc}(0)=1- (T/T_{c})^n$
 with $\sigma_{sc}(0)$ as a free parameter.\cite{Zimmermann95}  The
value of $T_c$ was taken from the magnetization measurements (see
Table~\ref{LambdaSSOIE}), $n$ was fitted as a common parameter for
all four data sets and was found to be $n=1.87(4)$, which is
typical for underdoped YBCO.\cite{Zimmermann95} Note, that the
exponent $n$ is isotope independent \cite{Hofer00,Khasanov03} and
that fits with the same $n$ for all data sets allow to increase
the accuracy in the determination of $\sigma_{sc}(0)$. The values
of $\sigma_{sc}(0)$ obtained from the fits are listed in
Table~\ref{LambdaSSOIE} and are in agreement with previous
results. \cite{Seaman90,Khasanov03} From the values of
$\sigma_{sc}(0)$ the relative isotope shift of the in-plane
penetration depth $\Delta
\lambda_{ab}^{-2}(0)/\lambda_{ab}^{-2}(0) = [\sigma_{sc}(X)(0) -
\sigma_{sc}(^{16}$O$_{pac})(0)] /\sigma_{sc}(^{16}$O$_{pac})(0)$
and the OIE exponent $\beta_{\rm O} = - d\ln
\lambda_{ab}^{-2}(0)/d\ln M_{\rm O}$ were determined (see
Table~\ref{LambdaSSOIE}). Here $X$ denotes $^{16}$O$_{\rm
p}$$^{18}$O$_{\rm ac}$, $^{18}$O$_{\rm p}$$^{16}$O$_{\rm ac}$, and
$^{18}$O$_{\rm pac}$. Our results (summarized in
Table~\ref{LambdaSSOIE}) are in good agreement with previous
estimates of SOIE from magnetization experiments \cite{Zhao96} and
our previous $\mu$SR non site-selective OIE results.
\cite{Khasanov03}


\begin{table*}[htb]
 \caption[~]{Summary of the SOIE on $T_c$ and $\lambda_{ab}^{-2}(0)$  for
Y$_{0.6}$Pr$_{0.4}$Ba$_2$Cu$_3$O$_{7-\delta}$ extracted from the
experimental data (see text for an explanation). The values of
$\Delta T_c/T_c$, $\alpha_{\rm O}$, $\Delta \lambda_{ab}^{-2}(0)/
\lambda_{ab}^{-2}(0)$, and $\beta_{\rm O}$ are not corrected for
the uncomplete
oxygen-isotope exchange (see text).  \label{LambdaSSOIE}} %
\begin{center}
\begin{tabular}{lccccccccccccc} \hline\hline

\hline Sample  & $T_c$  & $\Delta T_c$ & $\frac{\Delta
T_c}{T_c}$&$\alpha_{\rm O}$& $\sigma_{sc}(0)$  & $\Delta
\sigma_{sc}(0)$ & $\frac{\Delta
\lambda_{ab}^{-2}(0)}{\lambda_{ab}^{-2}(0)}$&$\beta_{\rm O}$ \\
&(K)&(K)&(\%)&&($\mu s^{-1}$)&($\mu s^{-1}$)&(\%)&\\
\hline
$^{16}$O$_{\rm pac}$&44.62(12)&--&--&--&1.090(7)&--&--&-- \\
$^{16}$O$_{\rm p}$$^{18}$O$_{\rm ac}$&44.66(10)&0.04(16)&0.09(36)&-0.007(29)&1.070(9)&-0.020(11)&-1.8(1.0)&0.14(8) \\
$^{18}$O$_{\rm pac}$&42.96(12)&-1.66(17)&-3.72(40)& 0.298(32)&1.022(9)&-0.068(11)&-6.2(1.0)& 0.50(8)\\
$^{18}$O$_{\rm p}$$^{16}$O$_{\rm ac}$&43.17(11)&-1.45(16)&-3.25(36)&0.260(29)&1.018(7)&-0.072(10)&-6.6(0.9)&0.53(7)\\
\hline \hline
\end{tabular}
\end{center}
\end{table*}


%
\begin{figure}[htb]
\includegraphics[width=1.0\linewidth]{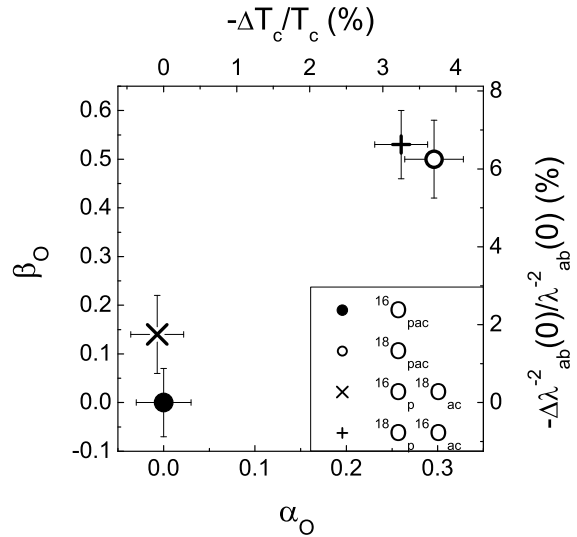}
\vspace{-0.5cm}
\caption{Plot of the  OIE  exponent $\beta_{\rm O}$ (-$\Delta
\lambda^{-2}_{ab}(0)/\lambda^{-2}_{ab}$(0)) versus the OIE
exponent $\alpha_{\rm O}$ (-$\Delta T_c/T_c$) for site-selective
Y$_{0.6}$Pr$_{0.4}$Ba$_2$Cu$_3$O$_{7-\delta}$ samples. The error
bars of the  ``trivial'' $^{16}$O$_{\rm pac}$ point (zero isotope
shift by definition) indicate the intrinsic uncertainty of the
experimental procedures on the estimation of $T_c$ and
$\lambda_{ab}$.}
\label{alpha-beta}
\end{figure}

In Fig.~\ref{alpha-beta},  we plot the OIE exponent $\beta_{\rm
O}$ ( $-\Delta\lambda^{-2}_{ab}(0)/\lambda^{-2}_{ab}(0)$) vs the
OIE exponent $\alpha_{\rm O}$ ($-\Delta T_c/T_c)$, in analogy with
Ref.~\onlinecite{Khasanov03}. Substantial  OIE on $T_c$ and
$\lambda^{-2}_{ab}(0)$ are observed for the completely $^{18}$O
substituted ($^{18}$O$_{\rm pac}$) and the site selective
$^{18}$O$_{\rm p}$$^{16}$O$_{\rm ac}$ samples, which are the same
within error bars. The other site-selective sample ($^{16}$O$_{\rm
p}$$^{18}$O$_{\rm ac}$) shows zero (within error bar) OIE on $T_c$
and a very small, but possibly non zero, OIE on
$\lambda^{-2}_{ab}(0)$. This effect very likely arises from the
small fraction of $^{18}$O that substituted $^{16}$O oxygen within
the CuO$_2$ planes, as seen in the Raman data (see
Fig.~\ref{Raman}). However, within the error bars, the OIE on
$\lambda_{ab}^{-2}(0)$ can still be considered the same for the
$^{16}$O$_{\rm pac}$ and $^{16}$O$_{\rm p}$$^{18}$O$_{\rm ac}$
samples.

In summary, we performed  $\mu$SR measurements of the
site-selective oxygen isotope effect on the in-plane magnetic
field penetration depth in underdoped polycrystalline
Y$_{0.6}$Pr$_{0.4}$Ba$_2$Cu$_3$O$_{7-\delta}$ samples. It was
found that the whole or nearly the whole effect on the transition
temperature $T_c$ and on the in-plane magnetic penetration depth
$\lambda_{ab}^{-2}$ (about 100~\% within error bars) comes from
the oxygen {\it within the superconducting CuO$_2$ planes} and not
from the apical and chain oxygen.  Noting that the lattice
parameters remain essentially unaffected by the isotope
substitution, \cite{Conder94,Raffa98} our results show the
existence of a strong coupling of the electronic subsystem to
phonon modes involving movements of the oxygen atoms in the
CuO$_2$ plane, while suggesting that modes involving apical and
chain oxygen are less strongly coupled to the electrons. This is
in agreement with recent inelastic neutron scattering
\cite{Mcqueeney99,Chung03} and photoemission
\cite{Bogdanov00,Lanzara01} studies, indicating  a strong
interaction between charge carriers and Cu-O bond-stretching-type
of phonons.

This work was partly performed at the Swiss Muon Source (S$\mu$S),
Paul Scherrer Institute (PSI, Switzerland). The authors are
grateful to A.~Amato and D.~Herlach for help during the $\mu$SR
measurements, D.~Di~Castro for helpful discussions, and L. Meier
for the measurements of the grain size distribution. This work was
supported by the Swiss National Science Foundation and by the NCCR
program MaNEP sponsored by the Swiss National Science Foundation.

\end{document}